\documentclass[aps,pre,reprint, amsmath, amssymb,superscriptaddress]{revtex4-1}

\usepackage{bm}
\usepackage[retainorgcmds]{IEEEtrantools}
\usepackage{graphicx}
\usepackage{mathrsfs}
\usepackage{amsmath}
\usepackage{amssymb,mathtools}
\usepackage{color}
\usepackage{amsfonts}
\usepackage{times,txfonts}
\usepackage{nicefrac}
\usepackage[colorlinks=true,linkcolor=blue,urlcolor=blue,citecolor=blue,pdfusetitle]{hyperref}
 \usepackage{soul}
\usepackage{cancel}

\definecolor{verde}{rgb}{.1,.6,.1}

\newcommand{\Cov}{\text{cov}}

\begin{document}

\title{Current fluctuations  in nonequilibrium discontinuous phase transitions}
\author{C. E. Fiore}
\email{fiorecarlos.cf@gmail.com}
\affiliation{Instituto de F\'{\i}sica da Universidade de S\~{a}o Paulo, 05314-970 S\~{a}o Paulo,  Brazil}
\author{Pedro E. Harunari}
\email{pedroharunari@gmail.com}
\affiliation{Instituto de F\'{\i}sica da Universidade de S\~{a}o Paulo, 05314-970 S\~{a}o Paulo,  Brazil}
\affiliation{Complex Systems and Statistical Mechanics, Physics and Materials Science Research Unit, University of Luxembourg, Luxembourg L-1511 G.D. Luxembourg}
\author{C. E. Fern\'andez Noa}

\author{Gabriel T. Landi}
\email{gtlandi@gmail.com}
\affiliation{Instituto de F\'{\i}sica da Universidade de S\~{a}o Paulo, 05314-970 S\~{a}o Paulo,  Brazil}

\date{\today} 
\begin{abstract}
Discontinuous phase transitions out of equilibrium can be characterized by the behavior of macroscopic stochastic currents.
But while much is known about the the average current, the situation is much less understood for higher statistics. In this paper, we address
 the consequences of the diverging metastability lifetime -- a hallmark of discontinuous transitions -- in the fluctuations of arbitrary thermodynamic currents, including the entropy production. 
In particular, we center our discussion on the \emph{conditional} statistics, given which phase the system is in.
We highlight the interplay between integration window and metastability lifetime, which is not manifested in the average current, but strongly influences the fluctuations. We introduce conditional currents and find, among other predictions, their connection to average and scaled variance through a finite-time version of Large Deviation Theory and a minimal model.
Our results are then further verified in two paradigmatic models of discontinuous transitions: 
Schl\"ogl's model of chemical reactions, and a  $12$-states Potts model subject to two baths at different temperatures. 
\end{abstract}

\maketitle

%
%
\section{Introduction}
%
%

In microscopic systems, currents of heat, work and entropy production must be treated as random variables, which fluctuate over different runs of an experiment~\cite{Seifert2012,RITORT2007528}. 
This represents a paradigm shift in thermodynamics, and has already led to fundamental advancements in the field, such as fluctuation theorems~\cite{Evans1993,Gallavotti1995,Jarzynski1997,Crooks1998,Esposito2009,Campisi2011} and, more
recently, the discovery of thermodynamic uncertainty relations~\cite{Barato2015,Pietzonka2015,Gingrich2016,Pietzonka2017a,Pietzonka2017}.
It also entails practical consequences, e.g.  in the design of Brownian engines~\cite{martinez2016brownian, blickle2012realization, proesmans2016brownian, quinto2014microscopic}, molecular motors~\cite{kinosita2000rotary,lieprl,liepre, lau2007nonequilibrium}, information-driven  devices~\cite{koski2014experimental,toyabe2010experimental}, and bacterial  baths~\cite{krishnamurthy2016micrometre}.
In these systems,  both the output power~\cite{Pietzonka2018,denzler2020power} and the efficiency~\cite{Verley2014a, verley2014universal, polettini2015efficiency,PhysRevResearch.2.032062} may fluctuate significantly, leading to possible violations of macroscopic predictions, such as the Carnot limit.~\cite{martinez2016brownian}.

A scenario of particular interest is that of non-equilibrium steady-states (NESSs), which occur when a system is placed in contact with multiple reservoirs at different temperatures $T_i$ and/or chemical potentials $\mu_i$.
NESSs are characterized by finite currents of energy and matter, and thus also a finite entropy production rate $\sigma_t$ ~\cite{Esposito2010d,VandenBroeck2010,Seifert2012,PhysRevE.82.021120,PhysRevE.91.042140}.
At the stochastic level, these become  fluctuating quantities, associated to a probability distribution.
Understanding the behavior of  said distributions constitutes a major area of research, as they form the  basis for extending the laws of the thermodynamics towards the microscale, 
providing insights in non-trivial properties of non-equilibrium physics.
Of particular interest is their behavior across non-equilibrium phase transitions~\cite{Marro1999}. 
Most of our understanding, however, is centered on the average current. 
For instance, the average entropy production rate has been found to be always finite around the transition point, with the first derivative either diverging, in continuous transitions~\cite{Tome2012,Shim2016,Crochik2005,Zhang2016,Noa2019,Herpich2019,PhysRevResearch.2.013136}, or presenting a jump in discontinuous ones~\cite{Zhang2016,Noa2019,Herpich2018}.
Conversely, the behavior of higher order statistics, such as the variance, is much less understood. 

Cumulants of thermodynamic currents are usually assessed  via numerical approaches, such as Monte Carlo simulations~\cite{Noa2019}, or large deviation theory (LDT)~\cite{Nguyen2020,Levitov1993,Flindt2009,Touchette2009,Esposito2009,Koza1999}.
In both cases, cumulants are computed from long-time sample averages, integrated over a time window $\tau$.
Ultimately, one is interested in taking $\tau \to \infty$, at least in principle. 
But in systems presenting discontinuous 
transitions this can become an issue, since the phase coexistence is characterized by states with very long metastability lifetimes $\tau_m$.
In fact, $\tau_m$  increases exponentially with the system volume $V$. 
As a consequence, the order of the limits  $\tau\to\infty$ and $V\to\infty$ becomes non-trivial~\cite{Baras1997}. 

In this paper we approach this issue by introducing the idea of conditional currents, given which phase the system is in. 
We focus, in particular, on the diffusion coefficient (scaled variance). 
We formulate a finite-time large deviation theory, which neatly highlights the non-trivial interplay between $\tau$ and $\tau_m$. 
This is then specialized to a minimal 2-state model, that is able to capture the key features of the problem and also provides useful predictions. 
These are then tested on two  paradigmatic examples of discontinuous transitions:
Schl\"ogl's model of chemical kinetics, and a  12-states Potts model subject to two baths at different temperatures. 

This paper is organized as follows: Sec.~\ref{sec:models} presents
the main concepts and assumptions considered.
The conditional large deviation theory is developed in Sec.~\ref{sec:LDT} and then specialized to a minimal model in Sec.~\ref{sec:minimal}. 
Applications are then considered in Sec.~\ref{sec:applications} and our conclusions are summarized in Sec.~\ref{sec:conc}.

%
%
\section{\label{sec:models}Models and assumptions}
%
%

We consider a stochastic system $X(t)$ undergoing Markovian evolution. 
For simplicity, we assume continuous-time and a discrete (possibly infinite) set of states $X(t) \in \mathcal{S}$. 
The system probability $p_x(t)$ is assumed to evolve according to the master equation~\cite{VanKampen2007}
\begin{equation}\label{M}
    \dot{p}_x(t) = \sum\limits_y \big\{W_{xy} p_y - W_{yx} p_x\big\} := \sum\limits_y \mathbb{W}_{xy} p_y,
\end{equation}
where $W_{xy} \equiv W_{y\to x}$ denotes the transition rates  from $y$ to $x$  and $\mathbb{W}_{xx} \equiv - \sum_{y\neq x} W_{yx}$.
The dynamics is taken to be ergodic, and such that $W_{xy}>0$ whenever $W_{yx}>0$, ensuring the system will relax to a unique steady state $p_x^*$.
In general, $p_x^*$ will be a non-equilibrium steady-state (NESS).

This NESS is also assumed to undergo a discontinuous transition by changing a certain control parameter $\lambda$ to a threshold value $\lambda_c$.
This means that in the vicinity of $\lambda_c$, there will exist a bistable region characterized by configurations with very long lifetimes.
The two phases are labeled as 0 (for $\lambda <  \lambda_c$) and 1 (for $\lambda > \lambda_c$). 
We monitor the phases by defining an indicator random variable $I_t = 0,1$ (henceforth called the \emph{phase indicator}), which specifies in which phase the system is at time $t$.
This can always be done by partitioning the set  of states $\mathcal{S}$ into two subsets, $\mathcal{S}_0$ and $\mathcal{S}_1$, representing each phase. 
The criteria for doing so is model dependent, and will be discussed further below. 
The probability of finding the system in phase 1, in the NESS, is then 
$q \equiv E(I_t) =  {\rm Pr}(I_t = 1)$.
We will also use the notation $q_1 = q$ and $q_0 = 1-q$, when convenient. 

The crucial aspect of discontinuous transitions is that, when the volume $V$ is large, transitions between coexisting phases become extremely rare. 
The system will thus be governed by two very distinct timescales, one describing fast relaxation \emph{within} each phase and another describing seldom transitions \emph{between} the phases.
The latter will be referred to as the metastability lifetime $\tau_m$, and usually grows exponentially with $V$ \cite{hanggi1984bistable}. 

We consider the consequences of this type of scaling to the behavior of a generic integrated thermodynamic current.
Given a certain time integration window $\tau$, such a current may be defined as~\cite{Gingrich2016}
\begin{equation}\label{current}
    \mathcal{J}_\tau = \int\limits_0^\tau dt \sum\limits_{y,z} d_{yz}  \delta_{X(t^-),y} \delta_{X(t^+),z},
\end{equation}
where $\delta_{ij}$ is the Kronecker delta, $X(t^{\pm})$ is the state of the system immediately before and after a transition and $d_{yz}$ is a function satisfying $d_{yz} = - d_{zy}$, which defines the current in question. 
In the limit $\tau\to \infty$, such a current will  behave according to a large deviation principle~\cite{Touchette2009}. 
But due to the sensitive interplay between $\tau$ and $\tau_m$, we will not assume $\tau\to \infty$, as is customary. 
Instead, we will analyze the behavior of $\mathcal{J}_\tau$ as a function of $\tau$. 
More specifically, our interest is in the regime where $\tau$ is large compared to the ``within-phase'' timescales, but not necessarily larger than the metastability lifetime $\tau_m$.
We will also focus on both the average  $J_\tau$, and diffusion coefficient (scaled variance) $D_\tau$, defined as
\begin{equation}\label{unconditional_quantities}
        J_\tau = E(\mathcal{J}_\tau)/\tau,
        \qquad 
        D_\tau = \Big(E(\mathcal{J}_\tau^2) - E(\mathcal{J}_\tau)^2\Big)/(2\tau).
\end{equation}
It turns out that $J_\tau \equiv J$ is  independent of $\tau$, irrespective of whether $\tau$ is large or not~\cite{Touchette2009}. 
Conversely, for $D_\tau$, this will be the case iff $\tau \gg \tau_m$.

The main feature we introduce in this paper is the notion of conditional currents, given which phase $i = 0,1$ the system is in. 
Inserting the identity $1 = (1-I_t) + I_t$ inside the integral~\eqref{current} allows us to define the current when the system is in phase 1 as 
\begin{equation}\label{J_stoch_1}
    \mathcal{J}_{\tau|1} = 
    \int\limits_0^\tau dt~I_{t^+} \sum\limits_{y,z} d_{yz}  \delta_{X(t^-),y} \delta_{X(t^+),z}.
\end{equation}
The current $\mathcal{J}_{\tau|0}$ is defined similarly, but with $1-I_t$ instead. 
There is an ambiguity here as to whether we use $I_{t^-}$ or $I_{t^+}$. 
But this only affects those jumps in which $I_{t^-} = 0(1)$ and $I_{t^+} = 1(0)$, which are extremely rare compared to all others. 
The total current~\eqref{current} is then recovered as
\begin{equation}\label{J_stochastic_split}
    \mathcal{J}_\tau =  \mathcal{J}_{\tau|0} +  \mathcal{J}_{\tau | 1}, 
\end{equation}
an identity which holds  at the stochastic level.

The conditional first moments are  defined as 
\begin{equation}\label{mu_i}
    \mu_i = \frac{E(\mathcal{J}_{\tau|i})}{\tau q_i},
\end{equation}
where the factor of $q_i$ in the denominator is placed to compensate for the varying times the system spends in each phase. The average current is thus decomposed as
\begin{equation}\label{aver}
    J = (1-q) \mu_0 + q \mu_1. 
\end{equation}
As with $J$, the conditional averages $\mu_i$ will be shown below to also be independent of $\tau$.

Similarly, we define conditional diffusion coefficients
\begin{equation}\label{D_i}
    D_{\tau|i} =  \frac{E(\mathcal{J}_{\tau|i}^2) - E(\mathcal{J}_{\tau|i})^2}{2\tau q_i},
\end{equation}
which represent the fluctuations of the system within each phase.
From Eq.~\eqref{J_stochastic_split}, we therefore see that the diffusion coefficient $D_\tau$ in Eq.~\eqref{unconditional_quantities} is split in three terms
\begin{equation}\label{diffusion_cond_uncond_relation}
    D_\tau = (1-q)D_{\tau|0} + q D_{\tau|1} + C_\tau, 
    \qquad C_\tau := \frac{1}{\tau} \Cov\big(\mathcal{J}_{\tau|0}, \mathcal{J}_{\tau|1}\big),
\end{equation}
where $\Cov(A,B) = E(AB) - E(A)E(B)$ is the covariance
between conditional currents $A$ and $B$, and is expected to be significant only in the vicinity of the transition point.

%
%
\section{\label{sec:LDT}Large deviation theory}
%
%

To shed light on the behaviour of conditional currents, we consider here a finite-time version of large deviation theory~\cite{Touchette2009}.
We being with the unconditional quantities, and then adapt our results to the conditional case. 
Let $G_\tau(\eta) = E(e^{\eta \mathcal{J}_\tau})$ denote the  moment generating function (MGF) associated to the current~\eqref{current}. 
Decomposing it as 
$G_\tau(\eta) = \sum_{x} E(e^{\eta \mathcal{J}_\tau} |X_\tau = x) p_x(\tau) = \sum_x G_x(\eta)$, we find that the entries $G_x(\eta)$ will evolve according to equation
\begin{equation}\label{MGF_evol}
    \frac{dG_x(\eta)}{d\tau} =  \sum\limits_{y} \mathbb{L}_{xy}(\eta) G_y(\eta),
\end{equation}
where the tilted operator $\mathbb{L}(\eta)$  depends on both the transition matrix $\mathbb{W}$ in Eq.~\eqref{M}, and the type of current in question, according to 
\begin{equation}
    \mathbb{L}(\eta)_{xy} = e^{\eta d_{xy}} \mathbb{W}_{xy},
\end{equation}
where, recall, $d_{xx} = 0$.
To evaluate $J$ and $D_\tau$, we only require the series expansion of $\mathbb{L}(\eta)$, which we write as
$\mathbb{L}(\eta) = \mathbb{W} + \eta L_1 + \eta^2 L_2$, for matrices $L_{1(2)}$ given by 
\begin{equation}\label{L1L2}
    (L_1)_{xy} = W_{xy} d_{xy}, 
    \qquad 
    (L_2)_{xy} = W_{xy} d_{xy}^2/2.
\end{equation}

\subsection{Unconditional cumulants}

We denote by $|\bm{p}\rangle$  the column vector whose entries are the steady-state distribution $p_x^*$, and $\langle \bm{1}|$ the row vector with all entries equal 1. 
Then, as discussed further in Appendix~\ref{app:ldt}, the first moment can be written, for arbitrary $\tau$,  as
\begin{equation}\label{LDT_J}
    J_\tau \equiv J = \langle \bm{1} |L_1 | \bm{p}\rangle,
\end{equation}
which is independent of $\tau$, as expected.
Conversely, the diffusion coefficient is written as 
\begin{IEEEeqnarray}{rCl}\label{LDT_D_int}
    D_\tau &=& \langle \bm{1}|L_2 |\bm{p}\rangle + \frac{1}{\tau} \int\limits_0^\tau d\tau' \int\limits_0^{\tau'} d\tau'' \langle \bm{1} |L_1 e^{\mathbb{W}(\tau'-\tau'')}L_1 |\bm{p}\rangle - \frac{J^2\tau}{2}.
    \IEEEeqnarraynumspace
\end{IEEEeqnarray}
We can also obtain a more explicit expression if we assume that $\mathbb{W}$ is diagonalizable, with eigenvalues $\lambda_i$, right eigenvectors $\mathbb{W}|\bm{x}_i\rangle = \lambda_i |\bm{x}_i\rangle$  and left eigenvectors $\langle \bm{y}_i | \mathbb{W} = \langle \bm{y}_i | \lambda_i$. 
Since the steady-state is unique, one eigenvalue must be zero, say  $\lambda_0 = 0$. 
The corresponding eigenvectors are then $|\bm{x}_0 \rangle = |\bm{p}\rangle$ and
$\langle \bm{y}_0 | = \langle \bm{1}|$.
Carrying out the integrals one then finds that 
\begin{equation}\label{LDT_D}
    D_\tau = \langle \bm{1} |L_2 |\bm{p}\rangle + \sum\limits_{i\neq 0} \langle \bm{1}|L_1 |\bm{x}_i\rangle\langle \bm{y}_i | L_1 |\bm{p}\rangle \left( \frac{e^{\lambda_i \tau} - 1 - \lambda_i \tau}{\lambda_i^2 \tau}\right),
\end{equation}
where we used the orthogonality relation $\langle \bm{1}|\bm{x_i}\rangle=0$, for  $i \neq 0$. 
This expression makes it clear that $D_\tau$ will depend sensibly on the interplay between $\tau$ and all eigenvalues $\lambda_i$ of $\mathbb{W}$.
If $\tau \gg 1/|\lambda_i|$, for all eigenvalues $\lambda_i \neq 0$, then the term $e^{\lambda_i \tau} - 1$ may be neglected, leading to the widely used expression from large deviation \begin{equation}\label{LDT_D_approx}
    D_\tau = \langle \bm{1} |L_2 |\bm{p}\rangle - \langle \bm{1}|L_1 \mathbb{W}^+ L_1 |\bm{p}\rangle,
\end{equation}
where $\mathbb{W}^+ = \sum_{i\neq 0} \lambda_i^{-1} |\bm{x}_i\rangle\langle \bm{y}_i|$ is the Moore-Penrose pseudoinverse of $\mathbb{W}$ (see Appendix~\ref{app:ldt} for more details).
Close to the transition point, there will appear a clear separation of time scales in the eigenvalues $\lambda_i$. 
At least one eigenvalue will be very small, of the order $\lambda_i \sim -1/\tau_m$, while all others will be much larger (describing the within-phase dynamics). 
If $\tau$ is large compared to these time scales, but not with respect to $\tau_m$, then 
the approximation taking Eq.~\eqref{LDT_D} to~\eqref{LDT_D_approx} will not hold true.
And since $\tau_m$ scales exponentially with the volume, as we approach the thermodynamic limit, larger and larger values of $\tau$ have to be considered.
This is a direct illustration of the non-commutativity of the limits $\tau \to \infty$ and $V\to\infty$.

\subsection{Conditional cumulants}

Eqs.~\eqref{LDT_J} and~\eqref{LDT_D_int} also apply to the conditional currents~\eqref{J_stoch_1}. 
One simply has to modify accordingly the tilted operator $\mathbb{L}(\eta)$ or, what is equivalent, the matrices $L_1$ and $L_2$ in Eq.~\eqref{L1L2}.
For each conditional current  $\mathcal{J}_{\tau|i}$, 
we  define a projection operator $\Pi^i$ such that $\Pi_{xy}^1 = \delta_{x,y} \sum_{z\in S_i} \delta_{y,z}$; i.e., which projects onto the states $\mathcal{S}_i$ associated to phase $i = 0,1$. 
The corresponding tilted operator will then be defined similarly, but with a current of the form $d_{xy}^i = d_{xy} \Pi_{yy}^i$, which means one should use instead  matrices $L_1 \Pi^i$ and $L_2 \Pi^i$.

Eq.~\eqref{LDT_J} then yields, taking also into account the factor $q_i$ in the denominator, 
\begin{equation}\label{LDT_mu_i}
    \mu_i = \frac{1}{q_i} \langle \bm{1}| L_1 \Pi^i |\bm{p}\rangle.
\end{equation}
Proceeding similarly with Eq.~\eqref{LDT_D_int}, we find 
\begin{widetext}
\begin{IEEEeqnarray}{rCl}\label{LDT_Di_int}
    D_{\tau|i} &=& \frac{\langle \bm{1} |L_2 \Pi^i |\bm{p}\rangle }{q_i} 
    +\frac{1}{\tau q_i} \int\limits_0^\tau d\tau' \int\limits_0^{\tau'} d\tau'' \langle \bm{1} | L_1 \Pi^i e^{\mathbb{W}(\tau'-\tau'')}L_1 \Pi^i|\bm{p}\rangle
    - \frac{\mu_i^2 q_i\tau}{2}.
\end{IEEEeqnarray}
And to obtain the covariance in Eq.~\eqref{diffusion_cond_uncond_relation}, we simply subtract the combination $(1-q) D_{\tau | 0} + q D_{\tau |1}$  from $D_\tau$ in Eq.~\eqref{LDT_D_int}. 
Recalling that $\Pi^0 + \Pi^1 = 1$, this then yields
\begin{IEEEeqnarray}{rCl}\label{LDT_C_int}
    C_\tau &=& \frac{1}{\tau} \int\limits_0^\tau d\tau' \int\limits_0^{\tau'} d\tau'' \langle \bm{1} |  L_1 \Pi^0 e^{\mathbb{W}(\tau'-\tau'')}L_1 \Pi^1|\bm{p}\rangle
    + \frac{1}{\tau} \int\limits_0^\tau d\tau' \int\limits_0^{\tau'} d\tau'' \langle \bm{1} | L_1 \Pi^1 e^{\mathbb{W}(\tau'-\tau'')}L_1 \Pi^0|\bm{p}\rangle
    - q(1-q)\mu_0\mu_1 \tau.
\end{IEEEeqnarray}
\end{widetext}
Concerning the timescales of the discontinuous transition, we notice that all diffusion coefficients, $D_\tau$, $D_{\tau|i}$ and $C_\tau$, are subject to a similar dependence, which is ultimately associated with the matrix $e^{\mathbb{W}(\tau-\tau')}$. 
Thus, one expects that all quantities should scale similarly with $\tau$. 

\subsection{\label{ssec:cond_dynamics}Conditioning on the dynamics}

There is a subtle, but crucial difference between conditioning the currents and conditioning the \emph{dynamics}. 
Eq.~\eqref{J_stoch_1} is an instance of the former: the current is conditioned on which phase the system is in, but $X(t)$ is still free to jump from one phase to the other. 
Alternatively, one could  define a conditional dynamics, where the system is forced to remain only within a certain phase.
This could be accomplished, for instance, by splitting the transition matrix $\mathbb{W}$ in Eq.~\eqref{M} in blocks of the form 
\begin{equation}
    \mathbb{W} = \begin{pmatrix}
    \mathbb{W}_{00} & \mathbb{W}_{01} \\[0.2cm]
    \mathbb{W}_{10} & \mathbb{W}_{11}
    \end{pmatrix},
\end{equation}
referring to the two subsets $\mathcal{S}_0$ and $\mathcal{S}_1$ of each phase. 
A conditional \emph{dynamics}, given phase $i$, is one that is governed by the restricted matrix $\mathbb{W}_{ii}$ (with  appropriate adjustments at the boundaries to ensure that it remains a proper transition matrix). 

One can similarly adapt Eqs.~\eqref{LDT_J} and~\eqref{LDT_D_int} to  this case. 
Let $|\bm{p}^i\rangle$ denote the steady-state of $\mathbb{W}_{ii}$.
For large volumes, since the two phases will be well separated, this will be quite similar to $\frac{1}{q_i}\Pi^i |\bm{p}\rangle$.
Applying Eq.~\eqref{LDT_J} will then yields exactly the same first moment $\mu_i$  in Eq.~\eqref{LDT_mu_i}. 
Hence, \emph{as far as the first moments are concerned, the distinction between conditional currents and conditional dynamics is thus irrelevant.}

However, for the diffusion coefficients this is absolutely crucial.
The reason is associated with the matrix exponential $e^{\mathbb{W}(\tau'-\tau'')}$ in Eq.~\eqref{LDT_D_int}.
Conditioning on the dynamics would lead instead to a matrix
$e^{\mathbb{W}_{ii}(\tau'-\tau'')}$. 
Since $\mathbb{W}_{ii}$ is essentially $\Pi^i \mathbb{W} \Pi^i$ (except for small modifications at the boundaries), we therefore see that the problem amounts to the difference between  $\Pi^i e^{\mathbb{W}(\tau'-\tau'')} \Pi^i$ (conditioning on the currents) and  $e^{\Pi^i \mathbb{W} \Pi^i(\tau'-\tau'')}$ (conditioning on the dynamics).
The two objects are \emph{drastically} different. 
The diffusion coefficients obtained by conditioning the dynamics, which we shall henceforth refer to as $\gamma_{\tau|i}$, will thus fundamentally different from the diffusion coefficients $D_{\tau|i}$ in Eq.~\eqref{D_i}.

An intuitive argument as to why this is the case  goes as follows. 
The currents~\eqref{J_stoch_1} are integrated over a certain time interval $\tau$. 
Hence, its diffusion coefficient will depend on correlations between different instants of time, and these are dramatically affected by the long timescale $\tau_m$ introduced by the discontinuous transition.
In fact, let us define $Z_t = \sum_{y,z} d_{yz} \delta_{X(t^-), y} \delta_{X(t^+),z}$, so that Eq.~\eqref{J_stoch_1} can be written as 
\begin{equation}
    \mathcal{J}_{\tau|1} = \int\limits_0^\tau dt~I_t Z_t. 
\end{equation}
The corresponding second moment will thus be 
\begin{equation}\label{cond_2nd_moment_partial}
    E(\mathcal{J}_{\tau|1}^2) = \int\limits_0^\tau dt\int\limits_0^\tau dt'~ E(I_t I_{t'} Z_t Z_{t'}).
\end{equation}
It hence depends, among other things, on the correlations between $I_t$ and $I_{t'}$, which decays very slowly around the transition point. 
For instance, in the simplest case where one can assume a Markovian 2-state evolution for $I_t$ (as will in fact be considered further in Sec.~\ref{sec:minimal}), one has 
\begin{equation}\label{2_time_corr}
    C(t-t') = \Cov(I_t, I_{t'}) = q(1-q) e^{-(t-t')/\tau_m},
\end{equation}
which will thus decay very slowly in time. 
This means that  $D_{\tau|i}$ in Eq.~\eqref{D_i} will depend very sensibly on the interplay between $\tau$ and $\tau_m$. 
Conversely, the diffusion coefficients $\gamma_i$, for the conditional dynamics, will not.
And hence, even for moderately large $\tau$, one expects it to be $\tau$-independent.

%
%
\section{\label{sec:minimal}Minimal model}
%
%

\begin{figure}
    \centering
    \includegraphics[width=0.45\textwidth]{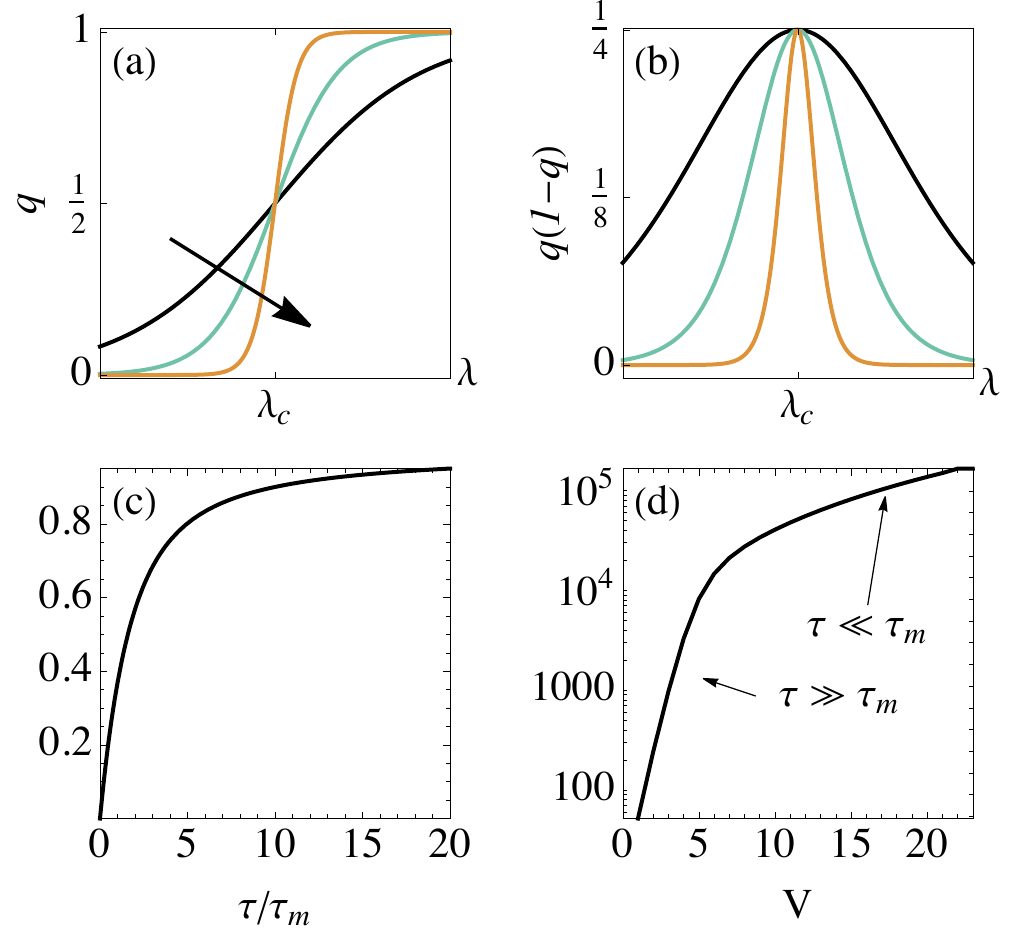}
    \caption{Predictions of the minimal model of discontinuous transitions.
    (a) The probability $q = (1+e^{-c V \Delta \lambda})^{-1}$ of finding the system in phase 1, for increasing volumes (depicted by the arrow). 
    (b) $q(1-q)$, which is non-negligible only in the vicinity of the transition point. 
    (c) The quantity 
    $(e^{- \tau/\tau_m} - 1 + \tau/\tau_m)/(\tau/\tau_m)$ appearing in Eq.~\eqref{LDT_D_final}. It tends to unity when $\tau\gg \tau_m$. 
    (d) Prototypical behavior of the diffusion coefficient~\eqref{LDT_D_final} as a function of volume, for a fixed $\tau$. 
    When $V$ is such that $\tau \gg \tau_m$, the diffusion coefficient grows exponentially with $V$. 
    But for a fixed $\tau$, as $V$ is increased, one must eventually cross the point $\tau \sim \tau_m$, after which the scaling becomes at most polynomial (due to the possible dependences of $\mu_i, D_i$ on $V$). 
    Parameters: $c_0 = c_a = c_b = \lambda_c = 1$, $\mu_0 = V/2$, $\mu_1 = 2V$, $\gamma_0 = \gamma_1 = V$. 
    }
    \label{fig:minimal}
\end{figure}

Many discontinuous non-equilibrium transitions can be approximated, for large volumes $V$, by a 2-state model~\cite{hanggi1984bistable}. 
That is, one reduces the dynamics essentially to the monitoring of the phase indicator $I_t$. 
In general, the dynamics of $I_t$ will be non-Markovian, as this would represent a hidden Markov chain. 
Instead, a minimal model is one where the dynamics of $I_t$ can be assumed to be Markovian, which is justified when $V$ is sufficiently large. 
In this case, instead of the full master equation~\eqref{M}, we may restrict the dynamics to 
\begin{equation}\label{M_2_level}
    \frac{d}{dt} q_i = \sum\limits_{j = 0,1} \mathcal{W}_{ij} q_j,
    \qquad 
    \mathcal{W} = \begin{pmatrix}
    -a & b \\[0.2cm]
    a & -b
    \end{pmatrix}.
\end{equation}
Here $a$ and $b$ represent the rates for the system to jump from phase $0\to1$ and $1\to 0$.
The steady-state  yields
$q \equiv q_1  =  E(I_t) = a/(a+b)$.
Moreover, the metastability lifetime in this case reads 
$\tau_m = 1/(a+b)$.
Finally, from~\eqref{M_2_level} one can compute the two-time correlation function, which is given in Eq.~\eqref{2_time_corr}.
And since $I_t$ can take on only two values, once $C(t-t')$ is known we can reconstruct the full joint distribution ${\rm Pr}(I_t = i, I_{t'} = i')$, for arbitrary times $t$, $t'$:
\begin{equation}\label{I_prob_corr}
    {\rm Pr}(I_t = i, I_{t'} = i') = \begin{cases}
    q^2 + C(t-t') & i=i'=1, \\[0.2cm]
    (1-q)^2 + C(t-t') & i=i'=0, \\[0.2cm]
    q(1-q) - C(t-t') & i\neq i'.
    \end{cases}
\end{equation}

The key feature of discontinuous transitions is the fact that transitions between phases are seldom when $V$ is large.
Close to $\lambda_c$, the transition rates $a$ and $b$ will usually behave, up to polynomial corrections, as
\begin{equation}\label{ab}
    a \sim e^{-V (c_0 - c_a \Delta \lambda)},
    \qquad 
    b \sim e^{- V(c_0 + c_b \Delta \lambda)},
\end{equation}
where $c_0,c_a,c_b >0$ are constants and $\Delta \lambda = \lambda - \lambda_c$. 
Note how the rates are  exponentially decreasing with  $V$. 
Transitions hence become rare when $V$ is large. 
From~\eqref{ab} we also get $\tau_m \sim e^{c_0 V}$, which is the aforementioned exponential dependence.
Finally, $q = (1+e^{-c V \Delta \lambda})^{-1}$, where $c=c_a+c_b>0$; hence $q$ changes  abruptly from 0 to 1 as $\lambda$ crosses $\lambda_c$, as illustrated in Fig.~\ref{fig:minimal}(a).
Since the conditional averages are weakly dependent on $\Delta \lambda$, from Eq.~\eqref{aver} we therefore see that $J$ should also change abruptly around $\lambda_c$, interpolating from $\mu_0$ to $\mu_1$. 

\subsection{Unconditional diffusion coefficient }

As shown in~\cite{Pietzonka2016a}, in this two-level model the tilted operator can be written, up to order $\lambda^2$, as 
\begin{IEEEeqnarray}{rCl}
\label{tilted_matrix}
    \mathbb{L}(\lambda) &=& \begin{pmatrix}
    -a + \lambda \mu_0 + \lambda^2 \gamma_0 & b \\[0.2cm]
    a & -b + \lambda \mu_1 + \lambda^2 \gamma_1
    \end{pmatrix}
    \\[0.2cm]
    &:=& \mathbb{W} + \lambda L_1 + \lambda^2 L_2.
\label{tilted_series}    
\end{IEEEeqnarray}
where $\gamma_i$ are the diffusion coefficients conditioned on the dynamics, not the currents (as introduced in Sec.~\ref{ssec:cond_dynamics}).

For the matrix $\mathbb{W}$ defined in Eq.~\eqref{M_2_level} we have $\lambda_1 = -1/\tau_m$, $|\bm{p}\rangle = (1-q,q)$,
$|\bm{x}_1 \rangle = (-1,1)$ and $|\bm{y}_1\rangle = (-q,1-q)$. 
Hence, using the explicit forms of $L_1$ and $L_2$ in Eq.~\eqref{tilted_series}, we get 
\begin{equation}\label{LDT_D_final}
    D_\tau = \gamma + q(1-q)(\mu_1-\mu_0)^2 ~\tau_m f(\tau/\tau_m),
\end{equation}
where $\gamma = (1-q)\gamma_0 + q \gamma_1$ is independent of $\tau$ and 
\begin{equation}\label{f_t}
    f(t) = (e^{- t} - 1 + t)/t.
\end{equation}
The interesting part is the last term in Eq.~\eqref{LDT_D_final}. 
First, it depends on $q(1-q)$, which is non-negligible only in the vicinity of the transition point (Fig.~\ref{fig:minimal}(b)). 
Second, it depends on the interplay between $\tau$ and $\tau_m$ through the function $f$, which is shown in Fig.~\ref{fig:minimal}(c).

When $\tau\ll \tau_m$ we get $f(\tau/\tau_m) \simeq \tau/2\tau_m$,
so that 
Eq.~\eqref{LDT_D_final} can be approximated to 
\begin{equation}\label{LDT_lim1}
    D_\tau \simeq \gamma + q(1-q)(\mu_1-\mu_0)^2 \tau/2, \qquad \tau \ll \tau_m,
\end{equation}
which is thus \emph{linear} in $\tau$.
Conversely, when $\tau \gg \tau_m$, we get 
\begin{equation}\label{LDT_lim2}
    D_\tau \simeq \gamma + q(1-q)(\mu_1-\mu_0)^2 ~\tau_m, 
    \qquad 
    \tau \gg \tau_m,
\end{equation}
which is independent of $\tau$, \emph{but linear in $\tau_m$}. 
Hence, when $V$ is large, this will become exponentially dominant. 
As a consequence, the large volume diffusion coefficient will actually become independent of the $\gamma_i$, and will instead be governed essentially by the mismatch in \emph{conditional averages} $(\mu_1-\mu_0)^2$, in agreement with previous studies on Schl\"ogl's model~\cite{Nguyen2020}. 

This offers another explicit illustration of the  order of  limits issue, which we depict graphically in Fig.~\ref{fig:minimal}(d):
For a given $\tau$, as we increase $V$ the diffusion coefficient will at first increase exponentially according to Eq.~\eqref{LDT_lim2}.
But if $\tau$ is fixed, then a point will always be reached around which $\tau \sim \tau_m$. 
And beyond this point, the scaling will be given by Eq.~\eqref{LDT_lim1}, which is at most polynomial in $V$ (due to a potential  polynomial volume dependence of $\mu_i,\gamma_i$).

Even though these results were developed for a 2-level model, they are still expected to hold for a broad class of discontinuous transitions. 
The reason is that, as discussed in Ref.~\cite{Vellela2009}, the eigenvalues and eigenvectors of the two-level transition matrix~\eqref{M_2_level} are connected to some of the eigenvalues and eigenvectors of the full matrix $\mathbb{W}$ in Eq.~\eqref{M}.
But, in addition, the full $\mathbb{W}$ will also have several other eigenvalues associated to the within-phase dynamics. 
Thus, the step from Eq.~\eqref{LDT_D} to~\eqref{LDT_D_final} only assumes that $\tau$ is much larger than all other $\lambda_i$, so that within-phase terms  can be neglected.

\subsection{Conditional diffusion coefficients} 

We can also use this minimal model to relate the diffusion coefficients $D_{\tau|i}$ in Eq.~\eqref{D_i} with the parameters $\mu_i, \gamma_i$. 
To do so, we use Eq.~\eqref{LDT_Di_int} with $\mathbb{W}$ now replaced by the two-state matrix $\mathcal{W}$ in Eq.~\eqref{M_2_level}. 
As a result, we find 
\begin{IEEEeqnarray}{rCl}
    D_{\tau|1} &=& \gamma_1 + \mu_1^2  (1-q)\tau_m f(\tau/\tau_m) ,
    \label{D1_minimal}
    \\[0.2cm]
    D_{\tau|0} &=& \gamma_0 + \mu_0^2q\tau_m f(\tau/\tau_m),
    \label{D0_minimal}
    \\[0.2cm]
    C_{\tau} &=& -2q(1-q)\mu_0 \mu_1   \tau_m f(\tau/\tau_m),
    \label{cov_minimal}
\end{IEEEeqnarray}
which can be combined together in the form~\eqref{diffusion_cond_uncond_relation}, to yield Eq.~\eqref{LDT_D_final}.
All conditional quantities are thus found to scale similarly with $\tau$, according to the function $f$ in Eq.~\eqref{f_t}. 
This allows us to conclude that even the conditional diffusion coefficients will be dominated by jumps between phases, and will be negligibly affected by the internal fluctuations within each phase.
We find this result both relevant and non-trivial.

It is also interesting to notice how the sign of the covariance~\eqref{cov_minimal} depends only on the signs of $\mu_0$ and $\mu_1$. 
A positively correlated covariance means that fluctuations above (below)  average in one phase tend to lead to fluctuations above (below) the average in the other; and vice-versa for $C<0$. 
We see in Eq.~\eqref{cov_minimal} that the covariance will be negative whenever $\mu_0,\mu_1$ have the same sign. 

%
%
\section{\label{sec:applications}Applications}
%
%

Next we shall exemplify our main findings in two representative systems displaying 
discontinuous phase transitions: The second Schl\"ogl~\cite{Schlogl1972} and a 12-state Potts models connected to two baths at different temperatures. 
The former was recently analyzed in Ref.~\cite{Nguyen2020}. 
It represents an ideal laboratory for testing our main prescriptions, since it presents an exact solution. 
The Potts model, on the other hand, is defined in a regular lattice and exhibits a nonequilibrium phase transition
under a different mechanism.
Despite the absence of an exact solution, all main features about the phase transition and statistics about entropy production fluctuations are present.

%
%
\subsection{Schl\"ogl's model}
%
%

\begin{figure*}
    \includegraphics[width=.95\textwidth]{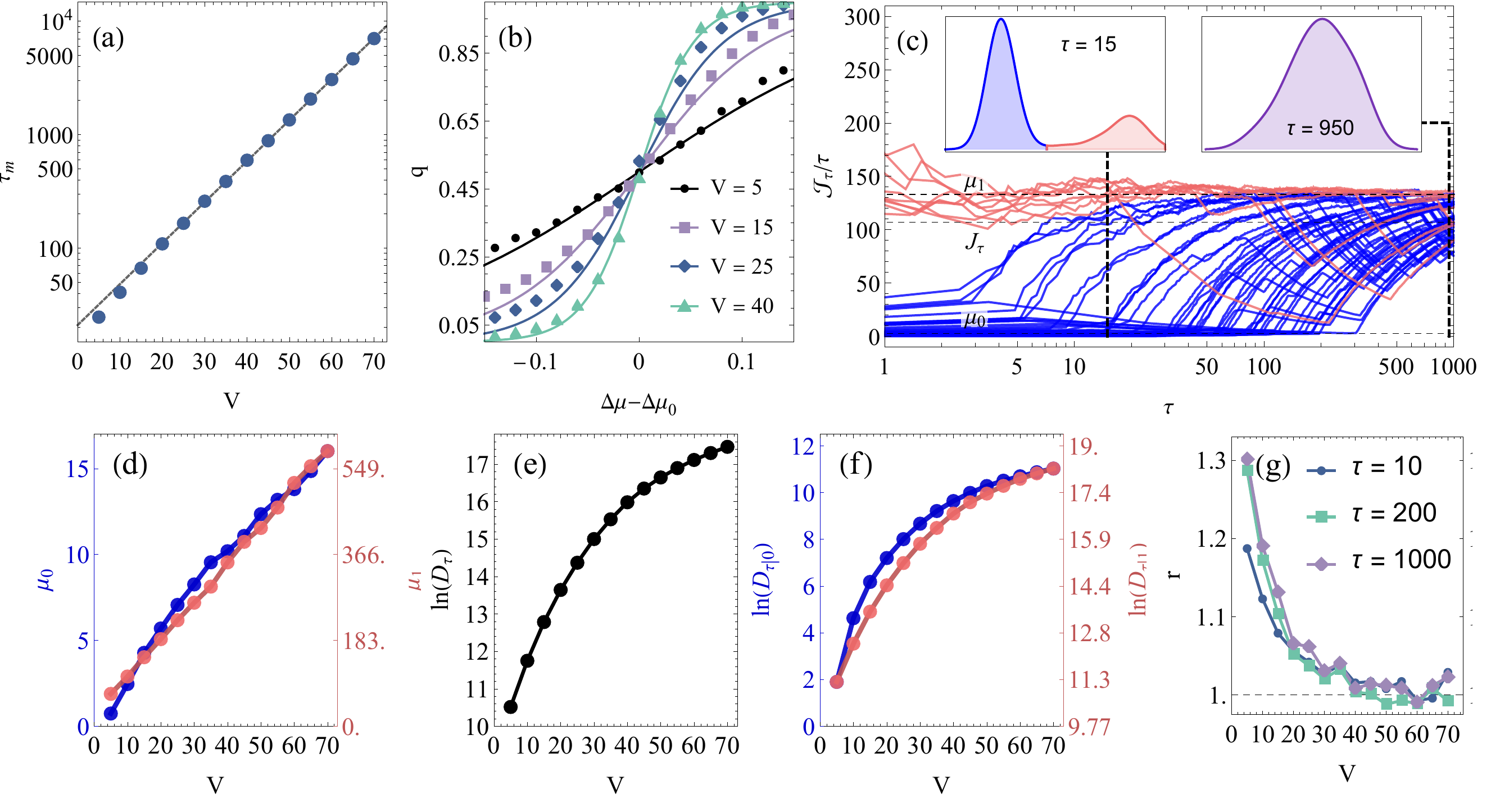}
    \caption{
    Conditional currents for Schl\"ogl's model, computed using the Gillespie algorithm.
    (a) $\tau_m$~vs.$V$.
    (b) $q$~vs.~$\Delta \mu - \Delta \mu_0$ for different values of $V$. Solid lines are a fit of $q = (1+e^{-c V (\Delta\mu-\Delta \mu_0)})^{-1}$. 
    (c) Stochastic trajectories of $\mathcal{J}_\tau/\tau$~vs.~$\tau$, starting either in phase 1 (red) or phase 0 (blue). The insets show the corresponding histograms at different times $\tau$. 
    (d)-(g) Mean and diffusion coefficient as a function of $V$, with $\tau = 10^3$ and $\Delta \mu$ fixed by setting $q = 1/2$. 
    (d) Conditional means $\mu_{0(1)}$ [Eq.~\eqref{mu_i}]. 
    (e) Diffusion coefficient $D_\tau$ [Eq.~\eqref{unconditional_quantities}].
    (f) Conditional diffusion coefficients $D_{\tau|i}$ [Eq.~\eqref{D_i}].
    (g) The ratio $r$ in Eq.~\eqref{ratio}, as a function of $V$, for different values of $\tau$. 
    Other parameters: \(a=k_1 = k_2 = k_{-2} = 1\) and \(b = 0.2\).
    }
    \label{fig:schlogl}
\end{figure*}

The second Schl\"ogl  model~\cite{Schlogl1972}
 describes a system with 3 chemical species, \(A\), \(B\) and \(X\), supporting two types of chemical reactions:
\begin{equation}\label{reactions}
2X+A \xrightleftharpoons[k_{-1}]{k_1} 3X,
\qquad
B \xrightleftharpoons[k_{-2}]{k_2} X.
\end{equation}
Here $k_{\pm 1}, k_{\pm2}$ are  kinetic constants that account, respectively, for catalytic, spontaneous creation and spontaneous annihilation of $X$. 
The concentrations of \(A\) and \(B\) are fixed at $a$ and $b$ due to the presence of chemostats. 
The dynamics of $p_n(t) = P(X(t)=n)$, for $n = 0,1,2,\ldots$, is then described by the master equation~\cite{Hanggi1984,Vellela2009}
\begin{equation}
    \dot{p}_n = f_{n-1} p_{n-1} + g_{n+1} p_{n+1} - (f_n + g_n) p_{n},
\end{equation}
where
\begin{IEEEeqnarray}{rl}
f_n &\coloneqq \frac{ak_1 n(n-1)}{V} + bk_2 V, \\[0.2cm]
g_n &\coloneqq \frac{k_{-1} n (n-1)(n-2)}{V^2} + k_{-2} n. 
\end{IEEEeqnarray}
The concentration 
$x(t) = X(t)/V$ presents a bistable behavior for large $V$~\cite{Vellela2009}, which is determined by the roots of the differential equation governing the deterministic behavior of \(x\) for large volumes
\begin{equation}
    \frac{dx}{dt} = ak_1 x^2 + bk_2 - k_{-1}x^3 -k_{-2}x = 0.
\end{equation}
The bistable region is defined as the  interval in the control parameters  for which this equation has three real roots, $x_0, x^*, x_1$. 
The first and last represent stable states for the most likely density within each phase, whereas $x^*$ is unstable and serves as the phase separator.
Hence, we define the phase-indicator in the Schl\"ogl's model as a random variable $I_t$ such that $I_t=1$ when $X(t) > V x^*$ and $0$ otherwise.]

For concreteness, we choose as thermodynamic current the entropy production $\mathcal{J}_\tau = \sigma_\tau$.
Whenever there is a transition, the net current~\eqref{current} changes by an increment $\delta \sigma_\tau$ defined according to the following rules: 
\begin{IEEEeqnarray}{rCLCLL}
\nonumber
2X + A &\xrightarrow{\;\;\;k_1\;\;\;}& 3X \qquad \qquad& \delta \sigma_\tau &=\mu_A ,
\\[0.2cm]
\nonumber
3X &\xrightarrow{\;\;\;k_{-1} \;\;\;}& 2X+A \qquad\qquad & \delta \sigma_\tau &=-\mu_A ,
\\[-0.2cm]
\label{transition_table}
&&&&& \\[-0.2cm]
\nonumber
X &\xrightarrow{\;\;\;k_{-2} \;\;\;}& B \qquad \qquad& \delta \sigma_\tau &=\mu_B ,
\\[0.2cm]
\nonumber
B &\xrightarrow{\;\;\;k_{2} \;\;\;}& X \qquad \qquad& \delta \sigma_\tau &=-\mu_B ,
\end{IEEEeqnarray}
where $\mu_A = \ln ak_1/k_{-1}$ and $\mu_B = \ln k_{-2}/bk_2$.

The model was simulated using the Gillespie algorithm.
We fix $ak_1 = k_{-2} = 1$, $bk_2 = 0.2$, and take as control parameter the chemical potential gradient $\Delta \mu =\mu_B-\mu_A = \ln \left[(k_{-2}a k_1)/(k_{-1} bk_2)\right]$. 
For these parameters, the phase coexistence point in the thermodynamic limit occurs at $\Delta \mu_0 \sim 3.047$~\cite{Nguyen2020}.
Figs.~\ref{fig:schlogl}(a) and (b) present a basic characterization of the steady-state. 
First, in Fig.~\ref{fig:schlogl}(a) we show the numerically computed metastability timescale $\tau_m$, as a function of the volume $V$, confirming the exponential dependence with $V$. 
This is obtained by collecting the mean first passage time 
$T_{x_i \to x^*}$
for the system to go from each stable point $x_{0(1)}$ to the unstable point, $x^*$. 
The rates $a$ and $b$ in Eq.~\eqref{M_2_level} are then given by 
$a = (2 T_{x_0\to x^*})^{-1}$ and $b = (2 T_{x_1\to x^*})^{-1}$~\cite{Gillespie1981}, from which we determine $\tau_m = 1/(a+b)$.
Second, Fig.~\ref{fig:schlogl}(b) characterizes the probability $q$ of finding the system in phase 1, as a function of $\Delta \mu -\Delta \mu_0$, for different values of $V$, where markers are simulation data and the curves are a fit of  $q = (1+e^{-c V (\Delta\mu-\Delta \mu_0)})^{-1}$; both agree very well for large volumes and/or small \(\Delta\mu - \Delta\mu_0\).
This is expected, since Schl\"ogl's model is known to have a well defined 2-state limit~\cite{Hanggi1984,Vellela2009} when $V$ is large.

Sample stochastic trajectories of the current $\mathcal{J}_\tau$ [Eq.~\eqref{current}] as a function of $\tau$ are shown in 
Fig.~\ref{fig:schlogl}(c), for fixed $\Delta \mu = 3.35$ and $V = 10$. 
Red and blue curves represent the situations where the system start in phases 1 and 0 respectively. 
For short $\tau$ the curves tend to remain well separated, so that $\mathcal{J}_\tau$ behaves as either $\mathcal{J}_{\tau|1}$ or $\mathcal{J}_{\tau|0}$. 
The corresponding statistics of $\mathcal{J}_\tau$, shown in the inset, would thus be a prototypical bimodal distribution. 
Conversely, when $\tau \gg \tau_m \sim 40$, transitions between the phases begin to occur, which cause the corresponding distribution to change to unimodal.

The conditional mean current and diffusion coefficients are shown in Fig.~\ref{fig:schlogl}(d)-(g).
For concreteness, we focus on the special point $q=1/2$; i.e., where the two phases are equally likely.
As this depends on $V$, for each volume we first fix $\Delta \mu$ as the point where $q=1/2$. 
This reduces the free parameters to $V$ and $\tau$ only. 
The conditional averages $\mu_{0(1)}$ as a function of the volume are shown in Fig.~\ref{fig:schlogl}(a). 
They are both found to be extensive in $V$, as expected; moreover, the activity in phase 1 is generally much larger, causing $\mu_1 \gg \mu_0$. 

Conversely, the diffusion coefficient $D_\tau$ (Fig.~\ref{fig:schlogl}(e)) and their conditional counterparts $D_{\tau|i}$ (Fig.~\ref{fig:schlogl}(f)) are both exponential in $V$, in line with previous studies~\cite{Nguyen2020}). 
For large volumes, these are also well described by the 
third term in Eq.~\eqref{LDT_D_final} (or~\eqref{D1_minimal}-\eqref{D0_minimal}).
We confirm this by plotting in Fig.~\ref{fig:schlogl}(g) the ratio 
\begin{equation}\label{ratio}
r = \frac{D_\tau}{q(1-q) (\mu_1-\mu_0)^2 \tau_m f(\tau/\tau_m)}, 
\end{equation}
where all quantities in the rhs are computed independently from the simulations. 
One can also consider similar definitions for $r_{0(1)}$. 
Since the $\gamma_i$ are at most polynomial in $V$, if this ratio tends to $r\to 1$ when $V$ is large, it serves as a confirmation that, for large $V$, the model effectively behaves as the 2-state minimal model of Sec.~\ref{sec:minimal}.
As is clear in Fig.~\ref{fig:schlogl}(g), this is indeed the case.

%
%
\subsection{$Q=12$-states Potts model}


As a second application, we study a $Q=12$ states Potts model coupled to two thermal baths at different temperatures.  
The model is defined in a regular 2D lattice with $V$ sites, where each site  $i$ assumes  one of $Q=12$ values $s_i = 1,\ldots,Q$ and interacts with its $z=4$ nearest neighbors, with energy 
${\cal H}(\bm{s})=-\sum_{i=1}^V\sum_{\delta=1}^z\delta_{s_i,s_{i+\delta}}$, 
where $\bm{s} = (s_1,\ldots,s_V)$. 
The equilibrium properties of this model have been studied extensively in~\cite{potts,kim19811,baxter1982magnetisation,fiorejcp2013,PhysRevLett.107.230601}. 
Here, we consider a non-equilibrium version where   the even and odd sites of the lattice are coupled  to thermal baths at temperatures $T_1$ and $T_1+\Delta T$ respectively, forming a checkerboard pattern. 
For concreteness, we fix $\Delta T = 0.9$.
This temperature gradient ensures a steady heat flux from one bath to the other, and hence a non-vanishing production of entropy \cite{Tome2012,Martynec_2020}. 

The model is simulated using standard Monte Carlo methods. 
The dynamics is assumed to be governed by Markovian single-site transitions  $s_i \to s_i'$, occurring with
rate  
$\omega_{s_i,s_i'}=\min\{1,\exp[-\Delta E_i/T_i]\}$,
where $\Delta E_i={\cal H}(\bm{s}')-{\cal H}(\bm{s})$
and $T_i$ is the temperature of site $i$.
For the current~\eqref{current}, we once again focus on the net entropy production rate to the environment which is characterized by increments $\Delta E_i/T_i$~\cite{Zhang2016,Martynec_2020}. 

\begin{figure}
    \centering
    \includegraphics[width=0.5\textwidth]{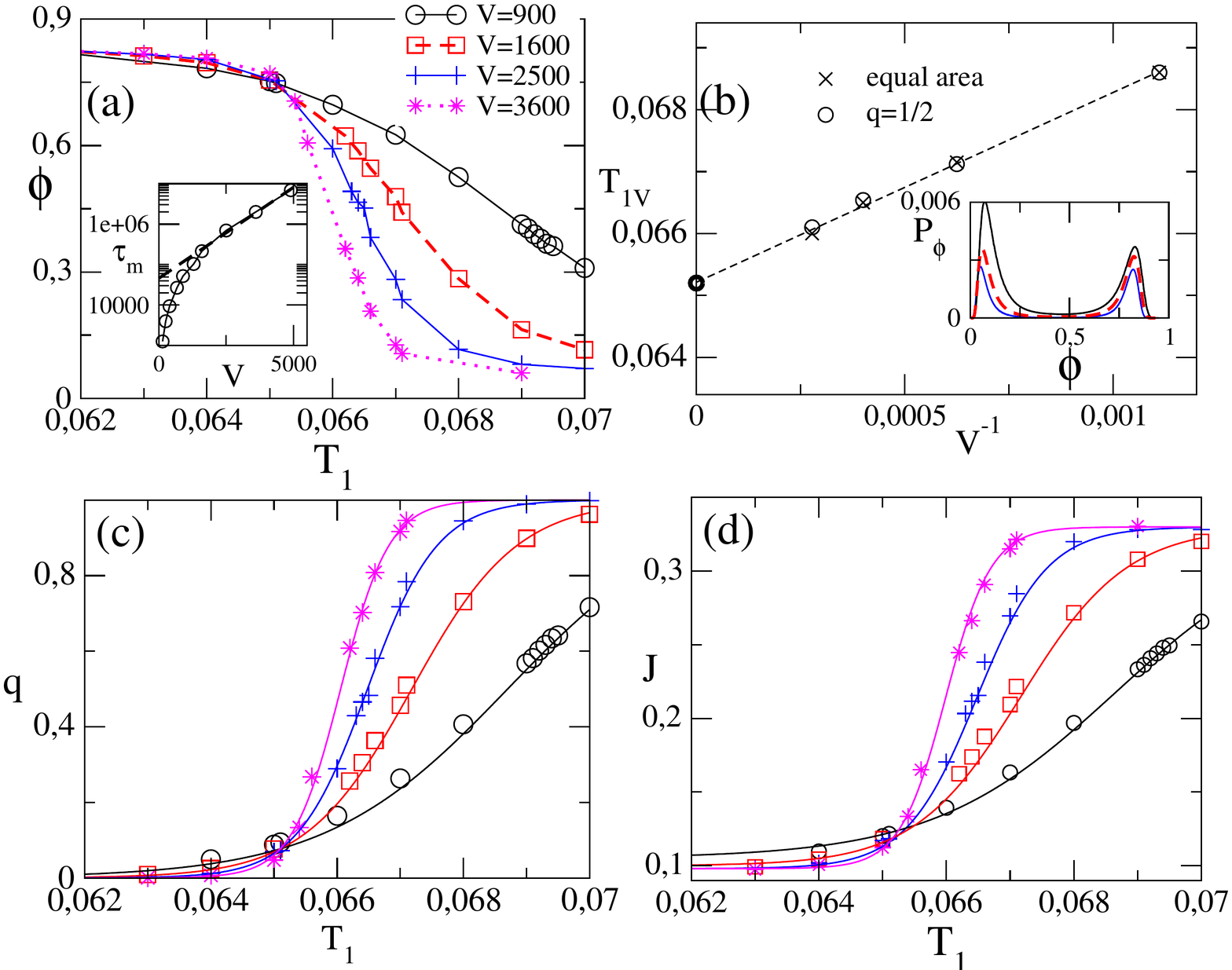}
    \caption{Characterization of the $Q=12$ Potts model in contact
    with two thermal baths of temperatures $T_1$ and $T_1+\Delta T$ (with fixed $\Delta T = 0.9$).  
    (a) Order parameter $\phi$~vs.~$T_1$ for different volumes $V$.
    Inset: metastability lifetime $\tau_m$~vs.~$V$. 
    (b) Finite-size analysis of the transition point $T_{1V}$~vs.~$V^{-1}$, yielding the asymptotic value $T_{01} = 0.0651(1)$. 
    Inset: distribution of $\phi$ at $T_{1V}$, for different volumes. 
    (c) Phase probability $q$~vs.~$T_1$, again for different volumes. 
    The continuous lines are fits of $q=[1 + Q e^{-Vc(T_1-T_{10})}]^{-1}$.
    (d) Average entropy production rate current $J$ [Eq.~\eqref{unconditional_quantities}], which follows closely the behavior of $q$. 
    }
    \label{fig:potts}
\end{figure}

As in the equilibrium version,
the phase transition is expected to be
discontinuous for $Q>4$.
Moreover, for $Q = 12$, the discontinuity is expected to become very sharp  for sufficient large $V$,  since it involves $Q$ distinct
ordered phases coexisting with a single disordered one.
The nonequilibrium phase transition can be quantified by the order-parameter $\phi={Q}[(\mathcal{N}_{max}/V)-1]/({Q}-1)$, where $\mathcal{N}_{max}={\rm max}\{\mathcal{N}_1,...,\mathcal{N}_{Q}\}$ is the maximum number of spins among all $Q$ configurations \cite{fiorejcp2013,challa}. 
Fig.~\ref{fig:potts}(a) shows results for $\phi$ as a function of $T_1$, for different lattice sizes $V$. 
The emergence of a discontinuous transition as $V$ increases is clearly visible. 
The inset in Fig.~\ref{fig:potts}(a) shows the metastability lifetime, which is again found to grow exponentially with $V$.

The sharp features of discontinuous phase transitions become  rounded at the vicinity of the coexistence point, due to finite size effects. 
To locate the transition point, we resort to the finite size scaling theory \cite{fiore2018}, establishing that  the ''pseudo-transition" point
$T_{1V}$, in which both phases have the same weight (equal-area order-parameter probability) reaches its asymptotic value $T_{10}$ according to the relation $T_{1V}-T_{10} \sim V^{-1}$. This is shown in Fig.~\ref{fig:potts}(b), from which we find  $T_{10}=0.0651(1)$.

A histogram of the order parameter $\phi$ is shown in the inset of Fig.~\ref{fig:potts}(b). It shows that  there is a clear separation between the two phases, allowing us to define a separator $\phi^* = 1/2$, such that the phase-indicator $I_t$ assumes the value $I_t=1$ when $\phi(t) >  \phi^*$.
The resulting phase probability $q$ is presented in Fig.~\ref{fig:potts}(c). 
As in the other models, it presents a sharp transition at $T_{10}=0.0651(1)$, and is well described by the expression $q=[1 + Q e^{-Vc(T_1-T_{10})}]^{-1}$.
Contrarily to Schl\"ogl's model, however, the curves for different volumes do not cross at $q = 1/2$, but instead at $q \simeq 1/13$. 
The unconditional current $J$ [Eq.~\eqref{unconditional_quantities}] is presented in Fig.~\ref{fig:potts}(d). 
As predicted by Eq.~\eqref{unconditional_quantities}, it follows very closely the behavior of $q$ [Fig.~\ref{fig:potts}(c)], smoothly interpolating between  $\mu_0$ and $\mu_1$.


\begin{figure}
    \centering
    \includegraphics[width=0.5\textwidth]{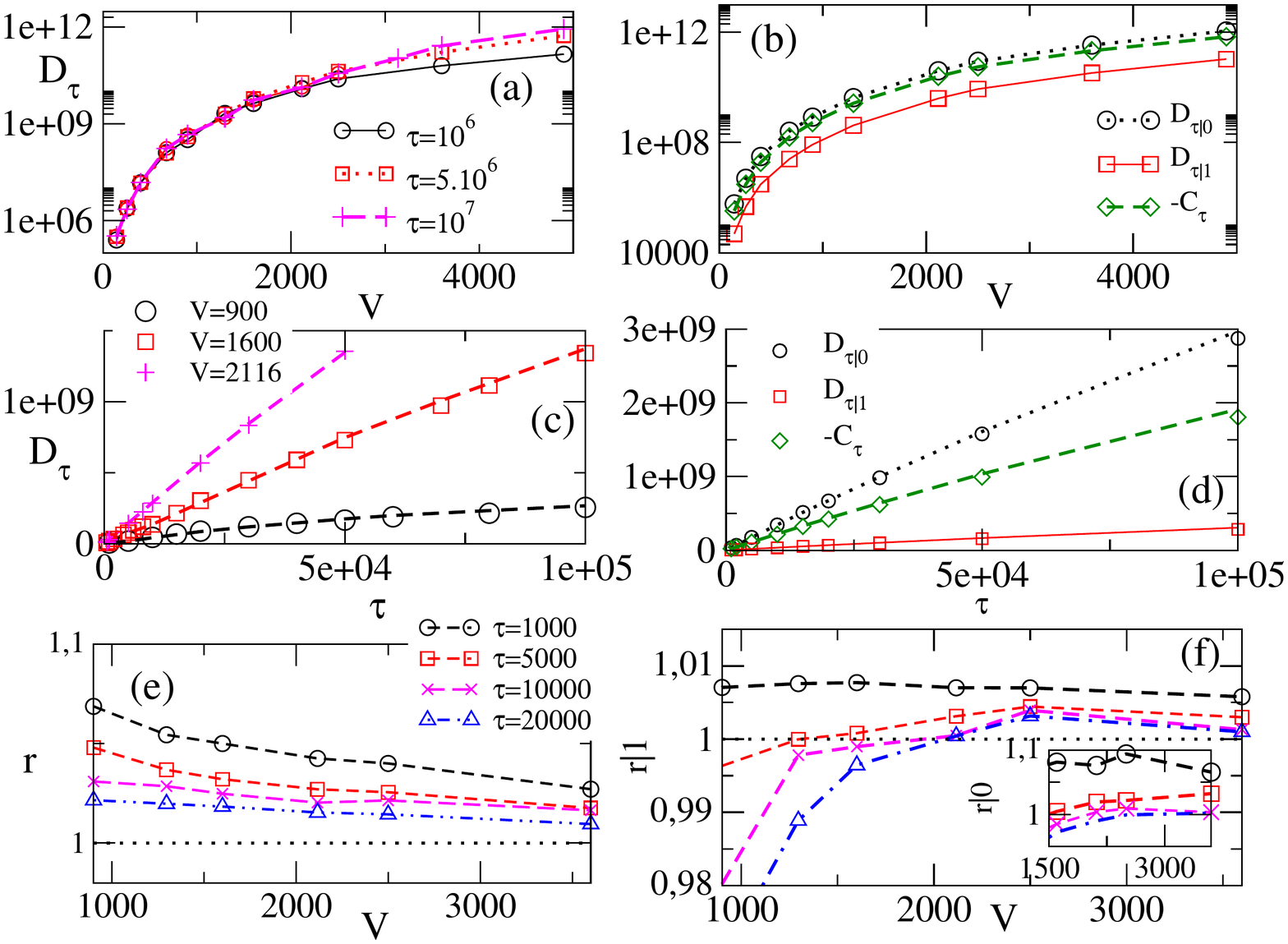}
    \caption{Unconditional and conditional diffusion coefficients for the $Q=12$ Potts model. 
    (a) $D_\tau$~vs.~$V$ for different values of $\tau$. 
    (b) $D_{\tau|i}$ and $C_\tau$~vs.~$V$ with $\tau = 5 \times 10^6$. 
    (c) $D_\tau$~vs.$\tau$ for different $V$. 
    (d) $D_{\tau|i}$ and $C_\tau$~vs.~$\tau$ for $V = 1600$. Continuous lines in (c) and (d) are the theoretical predictions from Eq.~(\ref{LDT_D_final}).
    (e) The ratio~\eqref{ratio} between $D_\tau$ and the predictions of the minimal model, Eq.~\eqref{LDT_D_final}, which tends to unity for large volumes. Curves are for different values of $\tau$. 
    (f) Same, but for $r|0$ (main plot) and $r|1$ (inset). 
    In all curves, for each $V$, we fix $T_1$ as the value $T_{1V}$ for which $q=1/2$. 
    Other details are as in Fig.~\ref{fig:potts}. 
    }
    \label{fig2q}
\end{figure}

\begin{figure}
    \centering
    \includegraphics[width=0.5\textwidth]{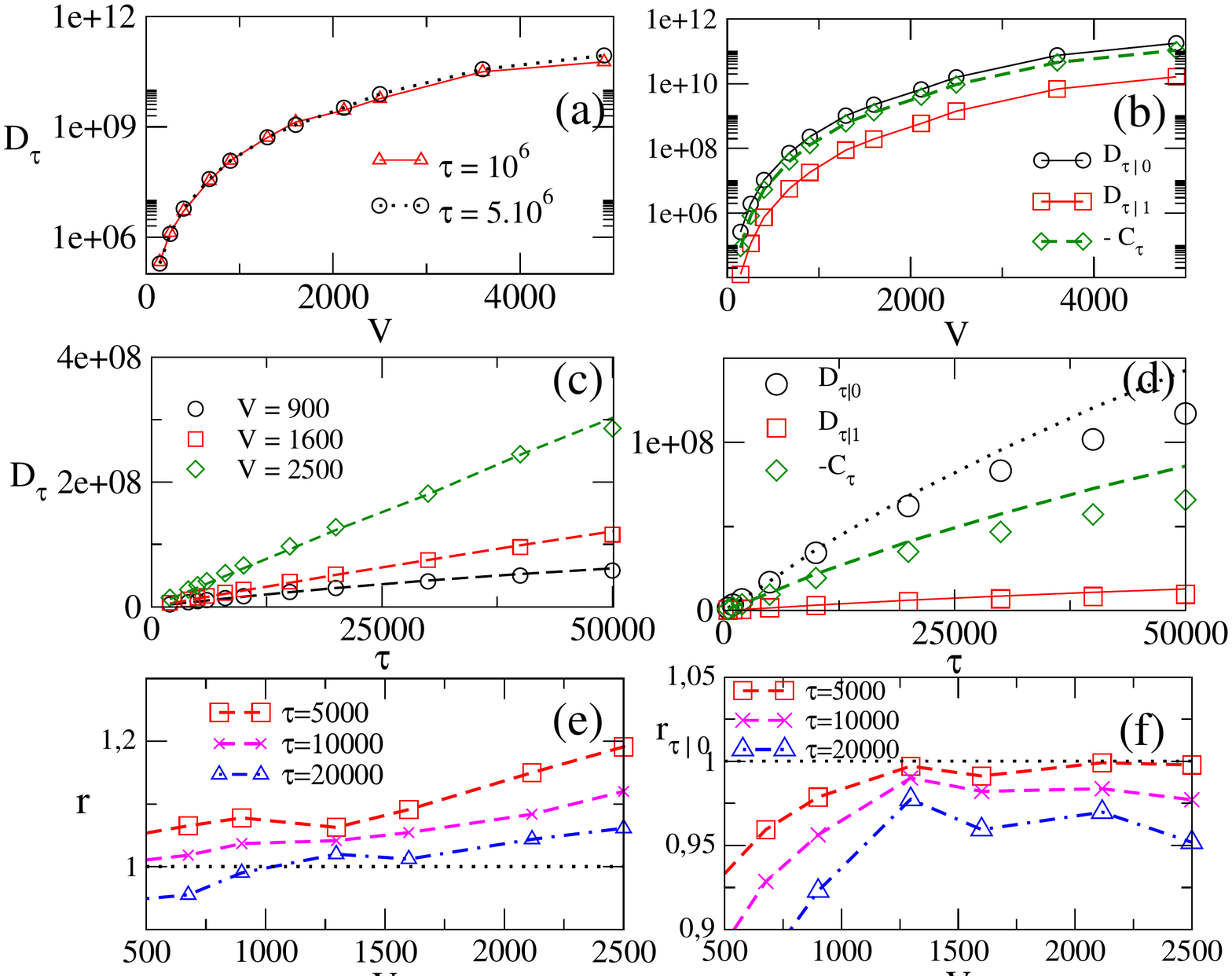}
    \caption{Same as Fig.~\ref{fig2q}, but fix $T_1$ fixed at the thermodynamic limit value $T_{10} = 0.0651(1)$. 
    }
    \label{fig2q2}
\end{figure}

We now turn to an analysis of the unconditional and conditional diffusion coefficients. 
The results are summarized in Fig.~\ref{fig2q}.
To reduce the number of free parameters, we proceed similarly to Schl\"ogl's model, and set, for each volume $V$, the temperature to $T_{1V}$ (i.e., so that $q=1/2$).
In Fig.~\ref{fig2q2} we repeat the same analysis, but fixing instead the temperature at $T_{10}$ (the thermodynamic limit transition point) for all $V$.
Similar findings are observed. 

The unconditional diffusion coefficient $D_\tau$ [Eq.~\eqref{unconditional_quantities}] is shown in Fig.~\ref{fig2q}(a) for different values of $\tau$. 
In agreement with the predictions of Eq.~\eqref{LDT_D_final}, for each $\tau$ the diffusion coefficient initially grows exponentially with $V$. But for a sufficiently large $V$, $\tau_m$ becomes comparable to $\tau$ and $D_\tau$ bends downwards.
This is exactly the behavior predicted by the minimal model [Fig.~\ref{fig:minimal}(d)].
The corresponding conditional diffusion coefficients are shown in Fig.~\ref{fig2q}(b). 
They follow a similar dependence on $V$ as $D_\tau$, which is in agreement with the expectations of Eqs.~\eqref{D1_minimal}-\eqref{cov_minimal}. 

The dependence of $D_\tau$, $D_{\tau|i}$ and $C_\tau$ as a function of $\tau$, for different $V$, are shown in Figs.~\ref{fig2q}(c),(d). 
In all cases, when $\tau$ is small the diffusion coefficients tend to be linear in $\tau$, in agreement with Eq.~\eqref{LDT_lim1}. 
If $V$ is not too large, then when $\tau$ becomes large one recovers instead a $\tau$-independent behavior, as predicted by Eq.~\eqref{LDT_lim2}. 
For large $V$ something similar is expected to occur, although it may require unrealistically large values of $\tau$. 

Finally, we study the ratio~\eqref{ratio}, between the actual diffusion coefficients and the predictions of the minimal model [Eq.~\eqref{LDT_D_final}]. 
The results, for both unconditional and conditional quantities, is shown in Fig.~\ref{fig2q}(e),(f).
In all cases, the plots clearly show that the ratio seems to tend to unity for sufficiently large $V$. 
This strongly indicates that the Potts model will also behave as an effective 2-state minimal model in the thermodynamic limit.

\section{\label{sec:conc}Conclusions}
The statistics of thermodynamic currents is
a fundamental issue in nonequilibrium thermodynamics, which has recently received significant interest. 
In this paper, we presented a simple and general description of the statistics of thermodynamic currents for systems displaying
discontinuous phase transitions. 
We introduced the idea of conditional statistics, accounting for the currents in each of the coexisting phases.
From large deviation theory, general relations for the unconditional and conditional cumulants of
a generic current were presented. We also proposed a minimal model, which captures all essential features of the problem. 
Our ideas were illustrated in two representative
systems: the exactly solvable 
Schl\"ogl's model of chemical reactions, and a  $Q$-states Potts model subject to two baths at different temperatures. 
In both cases, the results were found to follow very well the theoretical predictions of the minimal model, illustrating not only its reliability
but also the intricate role of distinct scaling times
and the volume.

As a final remark, we address some potential extensions of our work. It would be interesting to extend such approach to study the statistics of other quantities, such as the work. Another interesting point to be investigated concerns the use of our framework for tackling statistics of efficiency of thermal engines at the phase coexistence regimes.

\section*{Acknowledgments}
The authors acknowledge the financial support of the S\~ao Paulo Research Foundation FAPESP, under grants 
2018/02405-1,
2017/24567-0, 2020/03708-8, 
2019/14072-0.

\appendix
\section{\label{app:ldt} Large deviation theory results at arbitrary times}

In this appendix, we derive the expressions for
the first and second current moments 
from the  large deviation theory.
Unlike standard treatments, the main difference here is that we focus on finite integration times $\tau$. 
The starting point is Eq.~\eqref{MGF_evol}, describing the evolution of the entries $G_x(\eta)$ of the moment generating function (MGF).
Treating it as a vector $|\bm{G}(\eta)\rangle$ and  from
 its series expansion in powers of $\eta$, we have that
\begin{equation}
    |\bm{G}(\eta)\rangle = |\bm{p}\rangle + \eta |\bm{g}_1\rangle + \eta^2 |\bm{g}_2\rangle + \ldots,
\end{equation}
where $|\bm{p}\rangle$ is the steady-state of $\mathbb{W}$. 
Combining this with the series expansion of the tilted operator, $\mathbb{L}(\eta) = \mathbb{W} + \eta L_1 + \eta^2 L_2$, and collecting terms of the same order in $\eta$, we have the following system of equations
\begin{IEEEeqnarray}{rCl}
\label{eq_mom0}
    \frac{d}{d\tau} |\bm{p}\rangle &=& \mathbb{W}|\bm{p}\rangle, 
    \\[0.2cm]
    \frac{d}{d\tau} |\bm{g}_1\rangle &=& L_1 |\bm{p}\rangle + \mathbb{W}|\bm{g}_1\rangle,
    \label{eq_mom1}
    \\[0.2cm]
    \frac{d}{d\tau} |\bm{g}_2\rangle &=& L_2 |\bm{p}\rangle + L_1 |\bm{g}_1\rangle + \mathbb{W}|\bm{g}_2\rangle.
    \label{eq_mom2}
\end{IEEEeqnarray}
From these, the first and second moments are promptly obtained as
\begin{equation}\label{moments_g}
    E(\mathcal{J}_\tau) = \langle \bm{1} | \bm{g}_{1}\rangle,
    \qquad 
    E(\mathcal{J}_\tau^2) = 2 \langle \bm{1} | \bm{g}_{2}\rangle, 
\end{equation}
which follow from the definition of the MGF.
Eq.~\eqref{eq_mom0} is automatically satisfied in the steady-state. 
The solution of Eq.~\eqref{eq_mom1}, with $|\bm{g}_{1}(\tau=0)\rangle = 0$, is given by 
\begin{equation}\label{g1_sol}
    |\bm{g}_{1}(\tau)\rangle = \int\limits_0^\tau d\tau' e^{\mathbb{W}(\tau-\tau')} L_1 |\bm{p}\rangle.
\end{equation}
For concreteness, we assume $\mathbb{W}$ is diagonalizable as discussed above Eq.~\eqref{LDT_D}. 
We can then write
\begin{equation}\label{matrix_exp_expansion}
    e^{\mathbb{W}\tau} = |\bm{p}\rangle\langle \bm{1}| + \sum\limits_{i\neq 0} e^{\lambda_i \tau} |\bm{x}_i\rangle \langle \bm{y}_i|.
\end{equation}
The eigenvectors satisfy
$\langle \bm{1} | \bm{p}\rangle = \langle \bm{y}_i | \bm{x}_i \rangle = 1$ and $\langle \bm{1} |\bm{x}_i \rangle = \langle \bm{y}_i | \bm{p}\rangle = 0$.
Thus, plugging~\eqref{matrix_exp_expansion} in~\eqref{g1_sol}, we find 
\begin{equation}\label{g1_sol_detailed}
    |\bm{g}_1(\tau)\rangle = |\bm{p}\rangle\langle \bm{1}| L_1 |\bm{p}\rangle~\tau + \sum\limits_{i\neq 0} \frac{e^{\lambda_i \tau}-1}{\lambda_i} |\bm{x}_i \rangle \langle \bm{y}_i | L_1 |\bm{p}\rangle.
\end{equation}
To obtain the first moment we take the inner product  $\langle \bm{1}|\bm{g}_1\rangle$;  the second term vanishes and we are left with 
\begin{equation}
    E(\mathcal{J}_\tau) =\langle \bm{1}| L_1 |\bm{p}\rangle~\tau,
\end{equation}
which yields Eq.~\eqref{LDT_J}. 

Turning now to the second moment, the solution of Eq.~\eqref{eq_mom2} reads
\begin{IEEEeqnarray}{rCl}\label{g2_sol_tmp}
    |\bm{g}_2(\tau)\rangle &=& \int\limits_0^\tau d\tau' e^{\mathbb{W}(\tau-\tau')} (L_2 |\bm{p}\rangle + L_1 |\bm{g}_1(\tau')\rangle).
\end{IEEEeqnarray}
We are only interested in $\langle \bm{1}|\bm{g}_2\rangle$. 
Using Eq.~\eqref{matrix_exp_expansion}, together with the fact that $\langle \bm{1}| \bm{x}_i \rangle = 0$, we are then left only with
\begin{equation}
    \langle \bm{1} | \bm{g}_2(\tau) \rangle  = \int\limits_0^\tau d\tau' \Big\{\langle \bm{1} | L_2 |\bm{p}\rangle + \langle \bm{1} | L_1 |\bm{g}_1(\tau')\rangle\Big\}. 
\end{equation}
The first term is time-independent and hence will simply give a factor of $\tau$. 
In the second term we use Eq.~\eqref{g1_sol_detailed}, 
leading to 
\begin{equation}
    \langle \bm{1} | \bm{g}_2(\tau) \rangle 
    =
    \langle \bm{1} | L_2 |\bm{p}\rangle~\tau
    + 
    \int\limits_0^\tau d\tau' \int\limits_0^{\tau'} d\tau'' \langle \bm{1} |L_1 e^{\mathbb{W}(\tau'-\tau'')}L_1 |\bm{p}\rangle. 
\end{equation}
This, combined with the first moment squared, yields Eq.~\eqref{LDT_D_int}. 

To obtain the more explicit formula~\eqref{LDT_D}, we carry out the remaining integral, leading to 
\begin{IEEEeqnarray}{rCl} 
    \langle \bm{1} | \bm{g}_2(\tau) \rangle 
    &=&
        \langle \bm{1} | L_2 |\bm{p}\rangle~\tau
        + \langle \bm{1} | L_1 |\bm{p}\rangle^2 \frac{\tau^2}{2} 
        \\[0.2cm]
        &&+
        \sum\limits_{i\neq 0} \langle \bm{1} | L_1 |\bm{x}_i \rangle \langle \bm{y}_i  | L_1 |\bm{p}\rangle \left(\frac{e^{\lambda_i \tau} -1 - \lambda_i \tau}{\lambda_i^2} \right).
        \nonumber
\end{IEEEeqnarray}
The second term is  identified as the first moment squared. 
Hence, 
\begin{IEEEeqnarray}{rCl} 
    E(\mathcal{J}_\tau^2) - E(\mathcal{J}_\tau)^2 &=& 2  \langle \bm{1} | L_2 |\bm{p} \rangle~\tau
    \\[0.2cm] \nonumber
    &&+ 2 \sum\limits_{i\neq 0} \langle  \bm{1}|L_1 |\bm{x}_i \rangle\langle \bm{y}_i | L_1 |\bm{p}\rangle \left(\frac{e^{\lambda_i \tau} -1 - \lambda_i \tau}{\lambda_i^2} \right).
\end{IEEEeqnarray}
Dividing by $2\tau$ finally yields Eq.~\eqref{LDT_D}. 

As a final comment, concerning now the computation of Eq.~\eqref{LDT_D_approx}, which is valid when $\tau \gg \lambda_i$, it is convenient to express the solution in a way which is  independent of the full eigendecomposition of $\mathbb{W}$ (and hence more convenient for numerical computations). 
Let $|\bm{Q}_1\rangle$ denote the solution of the linear equation
\begin{equation}\label{linear_equation_ldt}
    \mathbb{W} |\bm{Q}_1\rangle = \Big(1 - |\bm{p}\rangle\langle \bm{1}|\Big) L_1 |\bm{p}\rangle. 
\end{equation}
This equation actually has an infinite number of solutions, which are of the form
\begin{equation}
    |\bm{Q}_1\rangle = \mathbb{W}^+ L_1 |\bm{p}\rangle + |\bm{p}\rangle\langle \bm{1}| \bm{w}\rangle, 
\end{equation}
for any vector $|\bm{w}\rangle$.
Here, recall, $\mathbb{W}^+$ is the Moore-Penrose pseudo-inverse of $\mathbb{W}$.
Projecting out the contributions from the subspace $|\bm{p}\rangle\langle \bm{1}|$, we see that 
\begin{equation}
    \Big(1 - |\bm{p}\rangle\langle \bm{1}|\Big)|\bm{Q}_1\rangle = \mathbb{W}^+ L_1 |\bm{p}\rangle. 
\end{equation}
Hence, Eq.~\eqref{LDT_D_approx} can be rewritten as 
\begin{equation}
    D_\tau = \langle \bm{1} |L_2 |\bm{p}\rangle - \langle \bm{1}|L_1 |\bm{Q}_1\rangle - \langle \bm{1}|L_1|\bm{p}\rangle \langle \bm{1}|\bm{Q}_1\rangle.
\end{equation}
This form of the diffusion coefficient is more familiar in the LDT literature, as compared with Eq.~\eqref{LDT_D_approx}.
It has the advantage that it requires solving a single linear equation~\eqref{linear_equation_ldt}, which is computationally much cheaper than fully diagonalizing $\mathbb{W}$.



\bibliography{library,lib2}
\end{document}